\documentclass{iaus}
\usepackage{graphicx,natbib}

\title[Eclipsing Binaries in Open Clusters]
{Eclipsing Binaries in Open Clusters}

\author[John Southworth]{John Southworth}

\affiliation{Department of Physics, University of Warwick, Coventry, CV4 7AL, UK \break
             email: jkt@astro.keele.ac.uk}

\pubyear{2007}
\volume{240}  
\pagerange{--}
\date{???? and in revised form ????}
\jname{Binary Stars as Critical Tools and Tests in Contemporary Astrophysics}
\editors{W.I. Hartkopf, P. Harmanec \& E.F. Guinan, eds.}

\begin{document}

\maketitle

\begin{abstract}
Detached eclipsing binaries are very useful objects for calibrating theoretical stellar models and checking their predictions. Detached eclipsing binaries in open clusters are particularly important because of the additional constraints on their age and chemical composition from their membership of the cluster. I compile a list containing absolute parameters of well-studied eclipsing binaries in open clusters, and present new observational data on the B-type systems V1481\,Cyg and V2263\,Cyg which are members of the young open cluster NGC\,7128.
\keywords{stars: binaries: eclipsing, stars: binaries: close, stars: distances, stars: evolution, stars: abundances, open clusters and associations: general}
\end{abstract}

\firstsection 

\section{The problem}

Theoretical stellar evolution models are of huge importance to stellar and galactic astrophysics because they are basically the only way of deriving the age, internal structure and composition of stars and galaxies from simple observational data (e.g.\ Meynet et al.\ 2005; Nordstr\"om et al.\ 2004). The accuracy of the results depends on the predictive power of the models, but this is severely limited by the parameteric treatment of several effects, including convective core overshooting ($\alpha_{\rm OV}$), the efficiency of convective energy transport ($\alpha_{\rm MLT}$), the effects of rotation, magnetic fields and mass loss, and the need to use theoretically-calculated opacities.

The main offender is $\alpha_{\rm OV}$, changes in which cause major modifications to the predicted lifetimes, chemical yields and luminosities of massive stars (e.g.\ Maeder \& Meynet 1989). This in turn has a large effect on the predicted properties and formation rates of evolved objects such as core-collapse supernovae (Eldridge \& Tout 2004) and black holes. Also, changes in $\alpha_{\rm MLT}$ can affect the derived ages of globular clusters by up to 2\,Gyr (Chaboyer 1995), whilst reasonable variation in $\alpha_{\rm OV}$ can alter the derived ages of open clusters by a factor of two (Bragaglia \& Tosi 2006). It is clear that this must be improved.

Theoretical models can be calibrated with observational constraints derived using detached eclipsing binaries (dEBs) (e.g.\ Pols et al.\ 1997; VandenBerg et al.\ 2006). The study of these objects allows us to accurately measure the masses, radii and $T_{\rm eff}$s of two stars of the same age and chemical composition (Andersen 1991). Observational errors in mass and radius can be below 1\% (e.g., Southworth et al.\ 2004e, 2005b). Accurate $T_{\rm eff}$s can also be found, allowing luminosities and distances to be directly measured.

dEBs have been used to constrain $\alpha_{\rm OV}$ (Andersen et al.\ 1990) and its dependence on mass (Ribas et al.\ 2000); $\alpha_{\rm MLT}$ was also studied by Ludwig \& Salaris (1999). However, the stength of these results has been compromised by the lack of information available for each object. For a dEB, the radii and effective temperatures are known for two specific masses. When comparing to theory, two of these four datapoints (usually the radii) are required to deduce the age and metal abundance of the system. The remaining two are generally not accurate enough to do anything more than indicate the helium abundance or, if you assume a normal helium abundance, perhaps the extent of convective overshooting. Further information is needed to derive accurate constraints on $\alpha_{\rm OV}$ or $\alpha_{\rm MLT}$. One possibility is to measure the surface metal abundances of the stars from high-dispersion spectroscopy, but this is rarely done. Another possibility is to study dEBs which display apsidal motion (rotation of the oritentation of an eccentric orbit due to tidal and relativistic effects; Claret \& Willems 2002) but these are less common and require observations over a long period of time.

\section{Solution: eclipsing binaries in open clusters}

\begin{table}
\caption{\label{EBOC} Parameters of well-studied dEBs in open clusters. Only systems with masses and radii known to within 5\% are included, sorted by decreasing primary star mass.}
\begin{tabular}{llcccl}\hline
dEB         & Cluster   & Masses (M$_\odot$) & Radii (R$_\odot$) & $T_{\rm eff}$s (K) & Reference \\ \hline
V1034 Sco   & NGC 6231  & 16.8$\pm$0.48 & 7.45$\pm$0.07 &  26200         & Bouzid et al.\ (2005) \\
            &           & 16.8$\pm$0.49 & 7.44$\pm$0.08 &  20350         & \\
V578 Mon    & NGC 2244  & 14.5$\pm$0.08 & 5.23$\pm$0.06 & 30000$\pm$500  & Hensberge et al.\ (2000) \\
            &           & 10.3$\pm$0.06 & 4.32$\pm$0.07 & 26400$\pm$400  & \\
V453 Cyg    & NGC 6871  & 14.4$\pm$0.20 & 11.1$\pm$0.13 & 26600$\pm$500  & Southworth et al.\ (2004b) \\
            &           & 8.55$\pm$0.06 & 5.49$\pm$0.06 & 25500$\pm$800  & \\
DW Car      & Cr.\ 228  & 11.2$\pm$0.10 & 4.54$\pm$0.05 & 27900$\pm$1000 & Southworth \& Clausen (2006a) \\
            &           & 10.5$\pm$0.12 & 4.28$\pm$0.05 & 26500$\pm$1000 & \\
V615 Per    & NGC 869   & 4.08$\pm$0.06 & 2.29$\pm$0.14 & 15000$\pm$500  & Southworth et al.\ (2004a) \\
            &           & 3.18$\pm$0.05 & 1.90$\pm$0.09 & 11000$\pm$1000 & \\
V906 Sco    & NGC 6475  & 3.25$\pm$0.07 & 3.52$\pm$0.03 & 10700$\pm$500  & Alencar et al.\ (1997) \\
            &           & 3.38$\pm$0.07 & 4.52$\pm$0.04 & 10400$\pm$500  & \\
GV Car      & NGC 3532  & 2.51$\pm$0.03 & 2.57$\pm$0.05 & 10200$\pm$500  & Southworth \& Clausen (2006b) \\
            &           & 1.54$\pm$0.02 & 1.43$\pm$0.06 &  8000$\pm$1000 & (provisional) \\
V618 Per    & NGC 869   & 2.33$\pm$0.03 & 1.64$\pm$0.07 & 11000$\pm$1000 & Southworth et al.\ (2004a) \\
            &           & 1.56$\pm$0.03 & 1.32$\pm$0.07 &  8000$\pm$1000 & \\
HD 23642    & Pleiades  & 2.19$\pm$0.02 & 1.83$\pm$0.03 &  9750$\pm$250  & Southworth et al.\ (2005a) \\
            &           & 1.55$\pm$0.02 & 1.55$\pm$0.04 &  7600$\pm$400  & \\
V392 Car    & NGC 2516  & 1.90$\pm$0.02 & 1.85$\pm$0.02 &  8850$\pm$200  & Debernardi \& North (2001) \\
            &           & 1.85$\pm$0.02 & 1.60$\pm$0.03 &  8650$\pm$200  & \\
V818 Tau    & Hyades    & 1.06$\pm$0.01 & 0.76$\pm$0.01 &  5530$\pm$100  & Torres \& Ribas (2002) \\
            &           & 0.90$\pm$0.02 & 0.77$\pm$0.01 &  4220$\pm$150  & \\
\hline \end{tabular} \end{table}

The study of eclipsing binaries in stellar clusters allows more constraints to be placed on theoretical models (e.g.\ Thompson et al.\ 2000; Lebreton et al.\ 2001). As the cluster stars, including both components of the dEB, were born in the same place at the same time, theoretical predictions must be able to match their observed properties for a single age and chemical composition. As a useful byproduct, the age and composition of the cluster can be derived from the theoretical models (e.g.\ Southworth et al.\ 2004ad) and its distance can be measured empirically from the properties of the dEB (e.g.\ Southworth et al.\ 2005ac). Fundamentally, this method requires a simultaneous fit of the morphology of the cluster in colour-magnitude diagrams and the properties of the dEB in mass-radius-temperature diagrams.

\begin{figure} \includegraphics[width=\textwidth,angle=0]{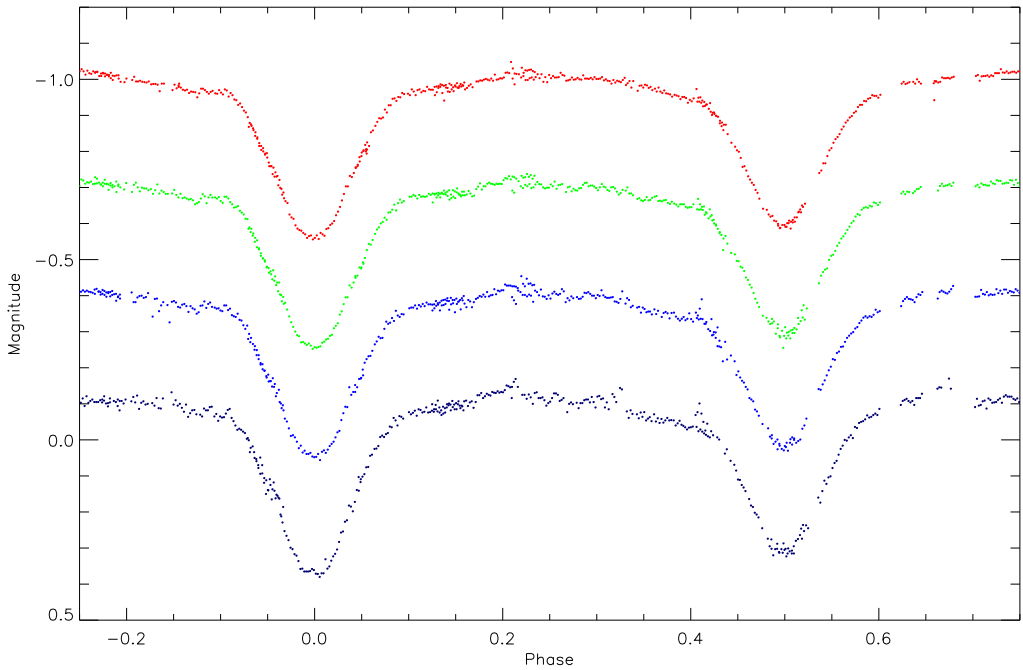}
\caption{\label{fig1} Preliminary (no debiassing or flat-fielding)
Str\"omgren $uvby$ light curves of the detached eclipsing binary
V1481\,Cyg in NGC\,7128 ($y$ is at the top and $u$ at the bottom).}
\end{figure}

\begin{figure} \includegraphics[width=\textwidth,angle=0]{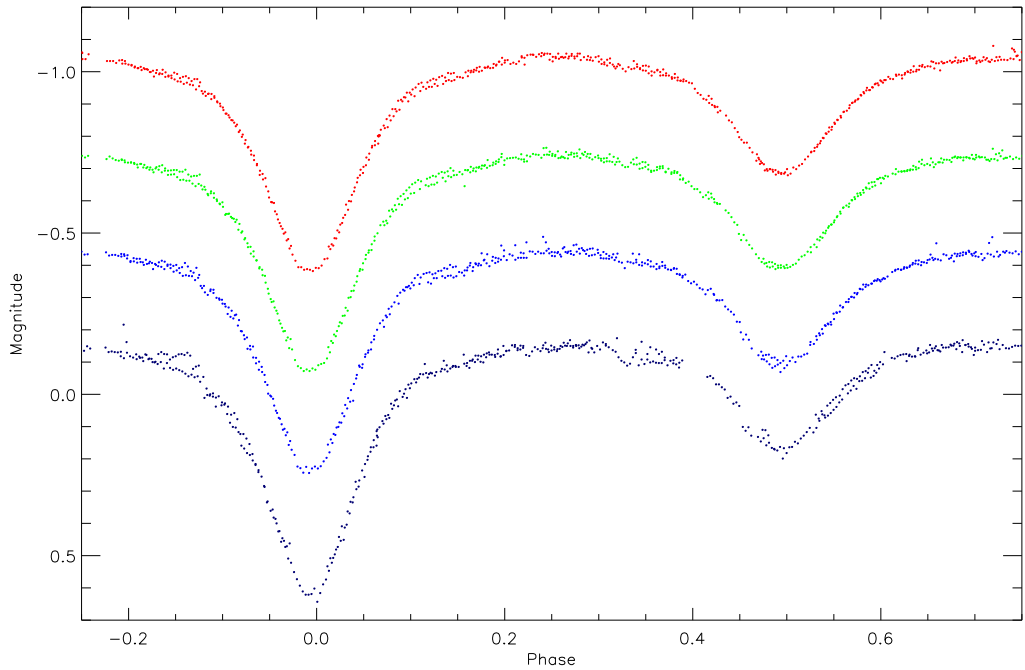}
\caption{\label{fig2} Preliminary (no debiassing or flat-fielding)
Str\"omgren $uvby$ light curve of the eclipsing binary V2263\,Cyg
in NGC\,7128 ($y$ is at the top and $u$ at the bottom). As it is
semi-detached it cannot be used to constrain single-star evolutionary
theory, but it can provide an accurate distance to the cluster.} \end{figure}

We are currently undertaking a research project to study dEBs which are members of open clusters with a range of ages and metallicities, in order to study how $\alpha_{\rm OV}$ and other theoretical model parameters vary with mass and composition. Whilst results so far have been very encouraging, it has still not been possible to match a well-studied dEB to a well-studied cluster; work to remedy this is in progress. One concept which has, however, come to light is that studying several dEBs in open cluster allows you to put four or more stars with the same age and composition but with differing masses into one mass-radius or mass-temperature plot (Southworth et al.\ 2004ac). This raises the possibility of obtaining ``standard sequences'' of eclipsing stars in one cluster with accurately known masses, radii, $T_{\rm eff}$s and surface gravities. To this end we have obtained photometry and spectroscopy of the eclipsing binaries V1481\,Cyg, V2261\,Cyg and V2263\,Cyg in the young open cluster NGC\,7128. Preliminary light curves for these are shown in Figs. \ref{fig1} and \ref{fig2}.


\end{document}